# Radiative heat transfer at nanoscale: experimental trends and challenges

Christophe Lucchesi[a†], Rodolphe Vaillon[b†], Pierre-Olivier Chapuis[a†*]

Energy transport theories are being revisited at the nanoscale, as macroscopic laws known since a century are broken at dimensions smaller than those associated with energy carriers. For thermal radiation, where the typical dimension is provided by Wien's wavelength, Planck's law and associated concepts describing surface-to-surface radiative transfer have to be replaced by a full electromagnetic framework capturing near-field radiative heat transfer (photon tunnelling between close bodies), interference effects and sub-wavelength thermal emission (emitting body of small size). It is only during the last decade that nanotechnology has allowed for many experimental verifications - with a recent boom - of the large increase of radiative heat transfer at nanoscale. In this minireview, we highlight the parameter space that has been investigated until now, showing that it is limited in terms of inter-body distance, temperature and object size, and provide clues about possible thermal-energy harvesting, sensing and management applications. We also provide an outlook on open topics, underlining some difficulties in applying single-wavelength approaches to broadband thermal emitters while acknowledging the promises of thermal nanophotonics and observing that molecular/chemical viewpoints have been hardly addressed.

## Context: revisiting thermal radiative energy transport and conversion at the nanoscale

The current energy challenge, which requires saving energy or energy harvesting in order to limit the greenhouse gas emissions, requires a fresh eye on many thermodynamic concepts. They date back to the XIX$^{th}$ century, when the laws of thermodynamics were established by famous names such as Carnot, Boltzmann, and many others. Heat transfer has been studied more accurately since that time. While heat conduction and convection are usually well understood (in particular because they are described by local partial differential equations), radiative heat transfer, i.e. the transfer of heat by photons, is often tackled in simplified ways. It is generally addressed either by considering surface-to-surface thermal radiation or radiation in semi-transparent/participating media. The first way considers thermal radiation only as a surface property, omitting that photon emission takes place in some matter volume. The second way, which is solved with the Radiative Transfer Equation, is more complex as non-local interaction is to be considered (volumes located very far from each other interact all the time). In both cases the wave nature of thermal radiation is largely neglected. Among the neglected phenomena, interference effects are of course to be mentioned, but other types of wave effects, such as tunnelling, are also out of the framework. Unfortunately, this is possible only at sizes much larger than the contributing wavelengths and therefore incorrect at nanoscale. In the last twenty years, a huge community dealing with nanophotonics, i.e. the engineering of matter to confer it particular properties for routing light and enhancing absorption/emission at particular wavelengths, has expanded. At the crossroad of these fields, i.e. energy management and nanophotonics, nanoscale thermal radiation effects are being studied and have the potential for many novel applications. In this context, this minireview aims at summarizing the types of configurations already tested experimentally and pointing to advances expected for applications of nanoscale radiative heat transfer in the near future, therefore complementing previous general reviews[1–3]. Thermal management, energy conversion and spectroscopy are among the key application fields.

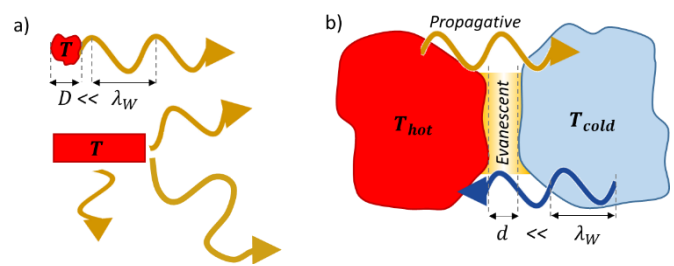

Fig. 1: Nanoscale radiative heat transfer. a) Sub-wavelength thermal emission from a body at a temperature $T$ having a size $D$ smaller than the characteristic wavelength of thermal radiation $\lambda_W$. Asymmetric bodies emit also differently along different directions. b) Near-field radiative heat transfer when the distance $d$ between two bodies having different temperatures is smaller than $\lambda_W$, taking place through evanescent wave tunnelling.

*Corresponding author: olivier.chapuis@insa-lyon.fr
[1] Univ Lyon, CNRS, INSA-Lyon, Université Claude Bernard Lyon 1, CETHIL UMR5008, F-69621 Villeurbanne, France
[2] IES, Univ Montpellier, CNRS, Montpellier, France
† Emails: christophe.lucchesi@insa-lyon.fr, rodolphe.vaillon@ies.univ-montp2.fr, olivier.chapuis@insa-lyon.fr.

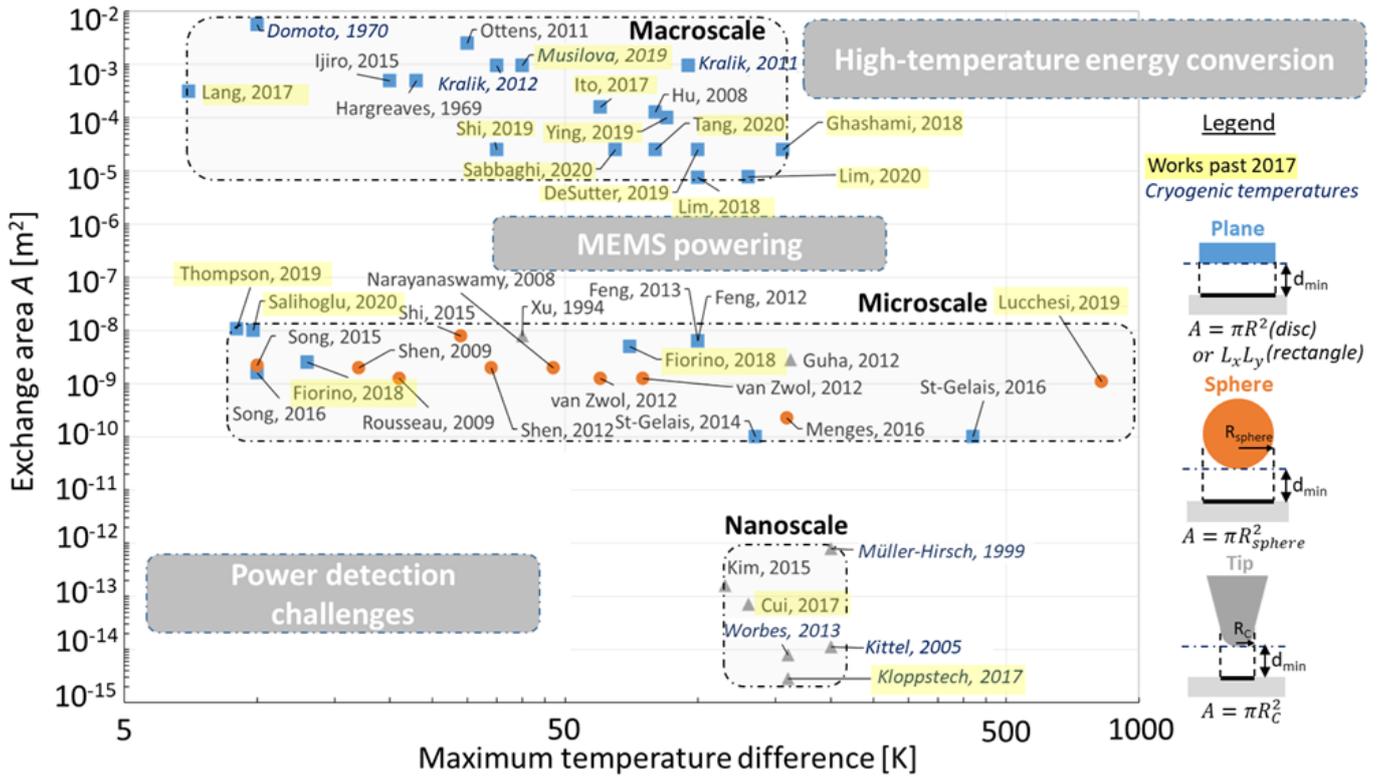

Fig. 2: Classification of the experiments involving near-field radiative heat transfer according to exchange area and temperature difference.

## Thermal radiation at nanoscale: key principles

The physics of thermal radiation is as follows: in matter, partially-charged particles, atoms or ions move, therefore generate electromagnetic waves which carry energy away. At equilibrium, such an energy loss is compensated by the local absorption of a similar amount of energy, which is possible due to the work of external electromagnetic fields on these partial charges. Since electromagnetic waves are screened when they travel in an absorbing medium, it can often be considered that only a skin layer at the boundaries of a body contributes to the associated radiative heat transfer, ensuing the surface-to-surface theory. At nanoscale, such condition is not easy to meet, and one could think that using the volume theory instead could be sufficient. An issue is that macroscopic theory cannot be applied when the considered sizes go down to the thermal radiation wavelengths, which lie in the micrometric range in general (recall that Wien's wavelength is given by $\lambda_W . T = 2890$ µm.K, where $T$ is temperature). For bodies of micrometric to sub-micrometric sizes, one has to consider *sub-wavelength thermal emission* (see Fig. 1(a)), and the emissivity concept associated with the surface theory breaks down (the simplest finite object, i.e. the sphere, has been addressed theoretically[4–6] while experimental evidences[7,8] have been provided, where the shape of the bodies also plays a role). In addition, the distance between bodies exchanging thermal radiation can also be sub-wavelength, when the bodies are interacting in their near field. This leads to *near-field radiative heat transfer* (see Fig. 1(b)) (note that both sub-$\lambda$ and near-field effects can happen at the same time). While the first type of effects has only been studied since very recently (see an outlook[9]), there is now a body of works for the second type, which was predicted a long time ago, in the frame of spatial (low temperature induces large near-field dimensions)[10]‡ and Casimir-related (fluctuational electrodynamics)[11] applications (see Reviews[12,13]). In particular, theory allows computing semi-analytically the energy exchange between shapes such as flat materials including multilayers[14], cylinders[15], cones[16], spheres[17–19] including nanoparticles[20–22] involving ellipsoidal[23,24] shapes and nanoparticle arrays, and gratings. Numerical methods have been developed in the last 5-10 years, allowing in principle to compute exactly radiative heat transfer between arbitrary shapes[25–27] provided one does not run up against the computing capability limits (some shapes can be numerically very-demanding). For smoothly-curved and close surfaces, the Proximity (Derjaguin) Approximation[28–31], which assimilates facing bodies to a sum of locally-planar and parallel surfaces, can be applied more easily. All these works have shown that the impact of the shape on near-field radiative heat transfer is strong. Finding the maximal radiative flux that can be transferred in the near field and ways to reach it is now an active field of research[32–34]. Finally, controlling thermal radiation properties by an external means is another active field of research. This can lead for instance to switches or logics-related devices (see initial proposals of rectifiers and transistors[35,36]). Such devices can be fully-thermal (this may require some heat input by another phenomenon than thermal radiation, for instance heat conduction) and based on phase change materials (VO$_2$ was used in experiments around room temperature[37–41]), or based on phenomena modifying optical properties, such as magnetic fields[42], strain, etc.. These devices could be useful in the medium to long term for thermal management in housing, in order to spare energy losses, or for computation (logics) with heat[43].

## Trends in experimental demonstrations

It is interesting to observe that the parameter space investigated experimentally so far is limited. The three key variables, temperature, power density and size, allow classifying the experiments performed up to now. First, one can observe that a vast majority of experiments were made with the body of lowest temperature at around 300 K (exceptions in italic in Fig. 2). Fig. 2 shows that only three types of geometries (plane-plane, microsphere-plane, and nanotip-plane) have been tested, associated each to a certain (projected) area of near-field radiative exchange. Large experiments are easier at low temperature, but up to ~cm$^2$ room-temperature devices have been demonstrated in the recent years, involving piezo-actuators, interferometric devices and capacitors. There seems to be a limit in the temperature difference $\Delta T$ reached so far, maybe due to the difficulty in managing large temperature variations in the surrounding elements in the setups. Note a recent experiment[44] for near-field thermophotovoltaic conversion could apparently test a larger $\Delta T$, but did not systematically investigate the temperature dependence. On the other hand, keeping microscale objects at larger temperature is easier, as illustrated by the microscale experiments. Finally, small temperature differences cannot be resolved by experiments with a small exchange area, which highlights the challenge of detecting low fluxes at the picowatt scale and below. The analysis of the transferred power is given in Fig. 3, where the thermal conductance (power per unit temperature) is represented vs the exchange area. To provide ideas of reference thermal conductances, diagonal lines indicate blackbody emissive power per kelvin ($4\sigma T^3$) at various typical temperatures (4, 77, 300 K) and a natural conductance associated with heat conduction at room temperature (ballistic conductance $<cv>/4$ that corresponds to the maximum heat transfer by conduction in a solid, where $<cv>$ is the spectral product of modal heat capacities and velocities). The figure shows that the maximal conductance detected in the near field indeed correlates approximately linearly with the area, and it can also be observed that near-field increases of two orders of magnitude at best over the far-field value have been demonstrated in many cases. Experiments with weak emitters such as metals can exhibit maximal conductances well below the blackbody limit (see e.g. the sub-120 K blue zone on the right). In almost all cases, near-field radiative conductances are far below conductive ones. Only experiments with scanning tunnelling microscopes (STM) lead to huge maximal thermal conductances just before electrical contact. Surprisingly, a large share of the experiments was conducted only with SiO$_2$ (known, in contrast to metals, for its high near-field contribution due to phonon polaritons and for coherence inducing narrow-band energy confinement[45]), which leaves the door open for studies of many other materials as will be discussed below.

## Applications enabled by the current state-of-the-art

The power exchanged depends clearly on size, and it is interesting to note that the three classes of experiments identified in Fig. 2 can be associated with various applications. **Large-area experiments**[46–58] address in particular the requirements for *near-field thermophotovoltaic* (TPV) conversion[44,59–62], where infrared (IR) energy emitted by a hot body is converted into electrical energy by a dedicated IR photovoltaic cell[63]. High-temperature thermal sources are expected to be used for energy harvesting. One possibility is first to concentrate solar light in order to heat up an emitter, in what is called *solar TPV*[64], and then benefit from the near-field increase to improve the device performances[65]. Near-field thermal radiation effects also play a critical role in *thermionic devices*[66], alone or in combination with TPV conversion[67,68]. Either used for *generating electricity* or *refrigerating*[69], the so-called near-field radiative thermoelectric energy converters[70] are all promising applications. However, scaling up laboratory proofs of concept system sizes[71] must chiefly address the challenge of building nanoscale gaps between macroscale planes with sufficiently large temperature differences, for example by using thermally low-conducting but mechanically stable spacers up to high temperature. Spacers made of polystyrene beads[48,51], pillars in SU-8 photoresist[56] or silica[49], or suspended devices[72,73] were used, leading to fixed distances for many current large-area experiments. Currently best gaps are of the order of few tens of nanometers for these experiments, with a recent claim of having reached the sub-10 nm region[74] with piezoelectric alignment. Extremely flat samples (roughness less than 1 nm) are required, which is possible thanks to silicon technology, without any surface bowing. The rapid deployment of autonomous devices and associated Internet of Things (IoT) calls also for miniaturized autonomous electrical sources. Harvesting small amounts of power from the environment could be performed by near-field TPV devices with smaller surface imprints, as shown by the **micrometric experiments**[19,37,39,40,72,73,75–86] reaching progressively the microwatt level with power densities close to 1 W.cm$^{-2}$ [62]. However, temperature differences required so far are still larger than the few-kelvin difference common situation widely encountered in real conditions. $\Delta T$s of 10-100 K may be observed in self-heating *micro-electromechanical systems* (MEMS), where such devices could be helpful. *Thermal management of electronics*, which can require the local cooling of hot spots of Integrated circuits (ICs), is probably also a natural field of application. Note also that micrometric-to-millimetric *bolometers* based on suspended membranes collecting radiation could be powered with similar devices (note the thickness-dependence of emissivity of such membranes[87,88]). **Small-area experiments**[89–94], essentially tip-based ones aiming at spatial investigation in the $xy$ plane, are currently limited to large temperature differences due to the lack of sensitivity and call for other types of method for detecting flux exchanges beyond the pW.K$^{-1}$ thermal conductance level around room temperature. Current methods involve devices based on the Seebeck effect (conversion of temperature difference into voltage by means of a thermocouple made of Au and Ni[89], Pt[90,91,93] or Cr[92,94])[95], on the temperature dependence of electrical conductivity (resistive thermometry)[39,62], and on bimaterial deflection involving an inhomogeneous temperature field in a cantilever[82–84,86,96]. At low temperature other thermometries are possible, for instance based on superconducting junctions[97]. Bimaterial deflection is in principle the most sensitive method (temperature variation down to 10$^{-5}$ K$^{-1}$ can be measured) but has been less popular recently, probably because it does not allow to easily scan in contact. Gold-coated SiN cantilevers were used for these experiments. Usually the tip is at a higher temperature than the sample, simply heated with the laser spot of the AFM, which limits the maximum achievable temperature. Higher temperature can be reached with self-heated resistive cantilevers[62]. Thermal sensitivity is especially important for spectroscopy, as only part of the power is located in a limited spectral range. Current *near-field thermal spectroscopy* is based on scattering the near-field energy to the far field, where it can be analysed[98–102]. Lock-in tip oscillation techniques may be needed as the spectral power is weak. However, the better sensitivity of IR photovoltaic cells/photodetectors[60,62] could lead to nanoscale thermal spectroscopic devices that would avoid the use of bulky Fourier-transform IR spectroscopes. Hyperspectral near-field thermal radiation imaging with such set-ups seems attractive. Current tip-based *imaging* devices without spectroscopes are based on flux measurements and were shown to be sensitive to monolayers on surfaces[91,103], while scattering techniques allow for detecting at different depths depending on the wavelength[98]. Finally, another fascinating field of study is the *transition between thermal radiation and heat conduction*, which could be investigated in the last nanometres before contact with AFM-based instruments[92] or even below with STM-based ones[90,93,94]. Note that non-local optics (permittivity with spatial dispersion $\epsilon(\omega,k)$) is expected to play a role at such gaps[104,105].

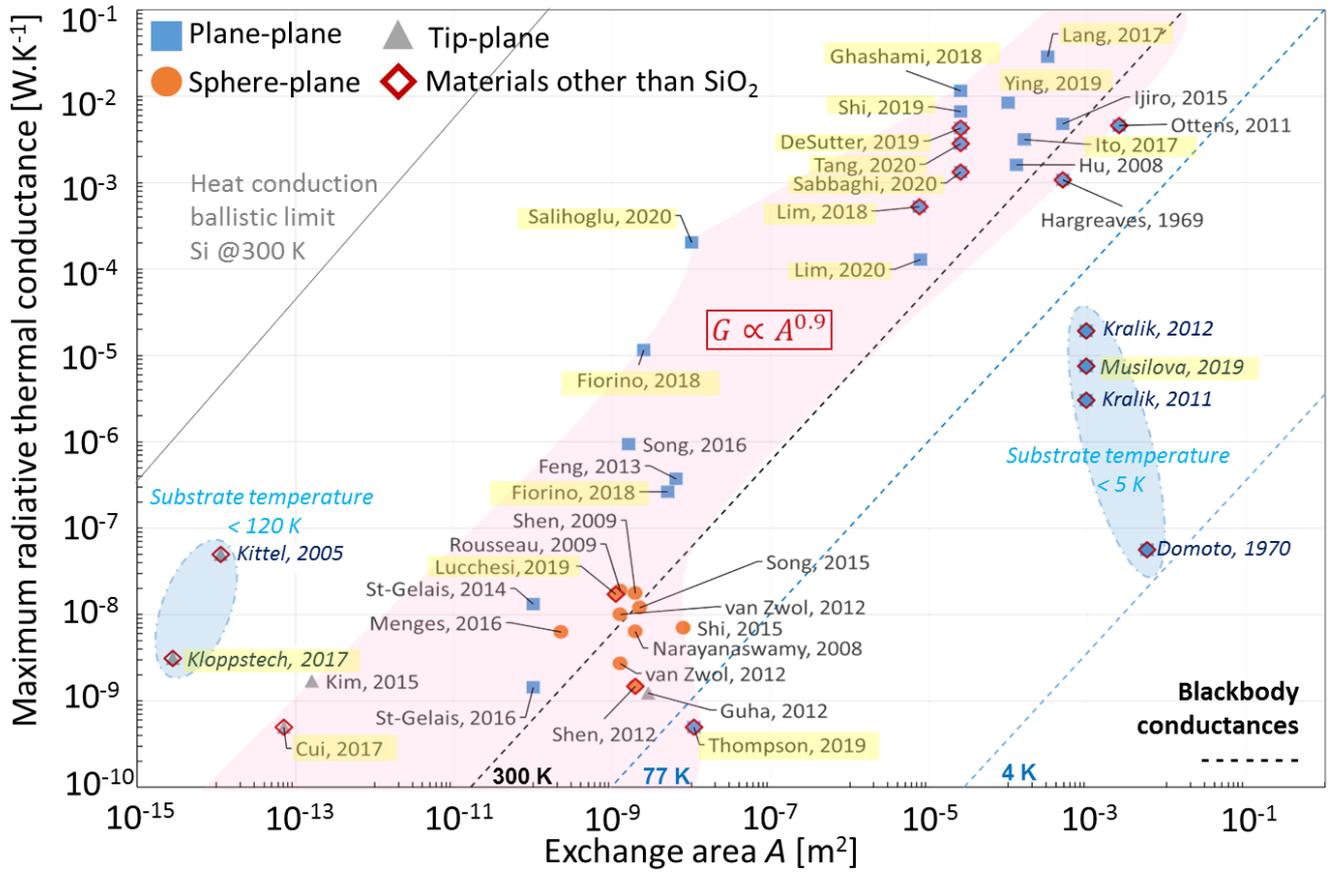

Fig. 3: Analysis of the maximal thermal conductance in near-field thermal radiation experiments as a function of exchange area. Blackbody conductances (dashed lines) are provided as references, as well as the maximal (ballistic) thermal conductance associated with heat conduction inside a solid (Si).

For many of the applications a key is to find a trade-off between the tailoring of the radiative spectrum and the need for large power transfer. While the radiative transfer can be constrained to one or more narrow bands (at a single wavelength maximal transfer of radiation between two bodies can be achieved, as the optical transmission coefficient can reach unity), for instance due to polariton resonance(s), this can lead to reducing the exchanged power to a minute value. The Bose-Einstein distributions associated with thermal photons unfortunately limits in principle the spectral power exchanged between bodies, however approaching the bodies in the near field and benefitting from the coupling between optical resonances allows to tunnel heat and increase the exchanged power. The complex electromagnetic field pattern showing up when more than two bodies are present was shown to allow increasing heat transfer[80].

## Outlooks related to thermal nanophotonics

Many of the experiments conducted so far involve at least one flat surface, while nanophotonics has been expanding since twenty years in various directions, in particular for focusing energy (see [106] for a review on the interest for far-field devices). There are avenues for spatially-resolved near-field thermal radiative devices, which would benefit from the knowledge gained in photonics. Here we discuss briefly few existing or possible ideas, related to **metamaterials**, **2D materials**, **nanoparticles/single objects**, and **scanning probes**, which may deserve original experimental investigations. The main parameters on which one would like to play are the *spectrum*, the *polarization* distribution, the *flux amplitude,* the *distance dependence* ($z$) and the *position dependence* ($xy$) of the flux. They may of course be entangled. Note that few works[91,96,103,107] have probed near-field thermal radiation of laterally nano/micro-architectured materials so far.

**Metamaterials** can cloak or focus energy, but they often work in a narrow band and it is not straightforward to translate the monochromatic results for a broadband spectrum. *Photonic crystals* allow to tailor the far-field spectrum and modify also the spectrum in the near-field with spectral gaps and resonant modes[57]. *Hyperbolic materials* possess dispersion relations leading to an enhanced heat transfer[96]. Fabry-Pérot like metal-insulator flat materials[54] can be included in this category. In addition, routing energy by means of *surface polaritons* has been predicted[108]: near-field transfer can be a way to excite and/or absorb them[109]. *Metasurfaces*[110,111] allow shaping wavefronts and could be useful in this respect, however broadband transfer may be difficult to achieve in practice. Manipulating the heat flow at sub-wavelength sizes in the $xy$ plane is appealing for the purpose of very large-scale integration (VLSI) of *IR sensors*, which are now constrained by the far-field diffraction limit associated with their wavelength: arrays of concentrating near-field devices operating at different wavelengths can be thought up.

**2D materials** controlled electrically such as gated graphene[57,85,112,113] or other single-atom thick materials like boron nitride, black phosphorus and transition-metal dichalcogenides (TMDC) appear interesting in order to control polaritons and therefore the intensity and spectrum of near-field radiative heat transfer. They can be combined with metamaterials such as hyperbolic materials (especially flat ones) as was shown in numerous theoretical proposals.

It is intriguing that applications of near-field radiative heat transfer with **nanoparticles** or **single objects** *(e.g.* nanoscale pads*)* standing on substrates have been difficult to envision until now, which is probably due to the low power they can transfer. However, their spectra are different from similar large-particle counterparts, and spectroscopic applications with intense fields are certainly possible[114]. *Arrays of nanoparticles*[109] can lead to photonic crystal properties, useful for tailoring emissive properties. The *magneto-optical* response of particles was shown to allow the flux to circulate in preferential paths (rotations)[42].

Such configurations could be probed by **scanning-probe tips** sensitive to magnetic fields, as U-shaped or MFM-based ones, in order to perform local characterization of the thermal fields, in particular *polarisation* effects. Taking advantage of *local field enhancement by resonances* on the sample surface is a way to increase the signal. This is reminiscent of molecular spectroscopy performed at a single wavelength, which could also be performed thermally in principle. We note that, while the interaction between nanoparticles was treated analytically[20] and that the atom/surface radiative exchange was studied in the frame of Bose-Einstein condensates[115], the radiative heat exchange between molecules and other bodies has not been treated extensively until now. Since single-molecule spectroscopy is a popular topic[116], there should be progresses in this direction in the coming years.

## Conclusions and prospects

The goal of this minireview has been to summarize the experimental efforts performed until now in the field of nanoscale radiative heat transfer. The analysis highlights various avenues, both for applications and fundamental science, which involve nanostructuring. In particular, it sheds light on the persisting challenge of building larger temperature differences (> 700 K) between macroscale areas separated by nanoscale gaps in energy conversion devices, or at the opposite on the need for improved thermal (sub-1 K) and transverse spatial resolution (sub-10 nm). It has also been shown that spectroscopic techniques avoiding scattering to the far field are proper alternative options. Many photonic techniques could be used for these purposes, however the strong entanglement between the *spectrum*, the *polarization* distribution, the *flux amplitude,* the *distance dependence* ($z$) and the *position dependence* ($xy$) of the flux is not easy to deal with and will require advanced engineering methods. At the crossroad of nanophotonics, optoelectronics, MEMS engineering, scanning probe microscopy and energy, nanoscale radiative heat transfer is now expected to spread among many applicative fields, including energy conversion, thermal management, or material analysis by means of imaging and spectroscopy techniques.

## Conflicts of interest

There are no conflicts to declare.

## Acknowledgements

The authors acknowledge discussions with many colleagues over the years, and critical reading by O. Merchiers. The authors acknowledge support from French ANR project DEMO-NFR-TPV

## Notes and references

‡ Note an incomplete treatment of evanescent waves at the time.

[1] J.C. Cuevas and F.J. García-Vidal, ACS Photonics (2018).

[2] B. Song, A. Fiorino, E. Meyhofer, and P. Reddy, AIP Adv. **5**, 53503 (2015).

[3] K. Parka and Z. Zhangb, Front. Heat Mass Transf. (2013).

[4] G.W. Kattawar and M. Eisner, Appl. Opt. **9**, 2685 (1970).

[5] M. Krüger, G. Bimonte, T. Emig, and M. Kardar, Phys. Rev. B **86**, 115423 (2012).

[6] K.L. Nguyen, O. Merchiers, and P.-O. Chapuis, Appl. Phys. Lett. **112**, 111906 (2018).

[7] D. Thompson, L. Zhu, R. Mittapally, S. Sadat, Z. Xing, P. McArdle, M.M. Qazilbash, P. Reddy, and E. Meyhofer, Nature **561**, 216 (2018).

[8] S. Shin, M. Elzouka, R. Prasher, and R. Chen, Nat. Commun. **10**, 1377 (2019).

[9] J.C. Cuevas, Nat. Commun. **10**, 3342 (2019).

[10] E.G. Cravalho, C.L. Tien, and R.P. Caren, J. Heat Transfer **89**, 351 (1967).

[11] D. Polder and M. Van Hove, Phys. Rev. B **4**, 3303 (1971).

[12] K. Joulain, J.-P. Mulet, F. Marquier, R. Carminati, and J.-J. Greffet, Surf. Sci. Rep. **57**, 59 (2005).

[13] A.I. Volokitin and B.N.J. Persson, Rev. Mod. Phys. **79**, 1291 (2007).

[14] M. Francoeur, M. Pinar Mengüç, and R. Vaillon, J. Quant. Spectrosc. Radiat. Transf. **110**, 2002 (2009).

[15] V.A. Golyk, M. Krüger, and M. Kardar, Phys. Rev. E **85**, 046603 (2012).

[16] A.P. McCauley, M.T.H. Reid, M. Krüger, and S.G. Johnson, Phys. Rev. B **85**, 165104 (2012).

[17] A. Narayanaswamy and G. Chen, Appl. Phys. Lett. **82**, 3544 (2003).

[18] K. Sasihithlu and A. Narayanaswamy, Opt. Express **22**, 14473 (2014).

[19] B. Guha, C. Otey, C.B. Poitras, S. Fan, and M. Lipson, Nano Lett. **12**, 4546 (2012).

[20] G. Domingues, S. Volz, K. Joulain, and J.J. Greffet, Phys. Rev. Lett. **94**, 085901 (2005).

[21] P.O. Chapuis, M. Laroche, S. Volz, and J.J. Greffet, Appl. Phys. Lett. **92**, 201906 (2008).

[22] A. Manjavacas and F.J. de Abajo, Phys. Rev. B **86**, 75466 (2012).

[23] S.-A. Biehs, O. Huth, F. Rüting, and M. Holthaus, J. Appl. Phys. **108**, 14312 (2010).

[24] R. Incardone, T. Emig, and M. Krüger, EPL (Europhysics Lett. **106**, 41001 (2014).

[25] M.T.H. Reid and S.G. Johnson, IEEE Trans. Antennas Propag. **63**, 3588 (2015).

[26] A.G. Polimeridis, M.T.H. Reid, W. Jin, S.G. Johnson, J.K. White, and A.W. Rodriguez, Phys. Rev. B **92**, 134202 (2015).

[27] S. Edalatpour and M. Francoeur, J. Quant. Spectrosc. Radiat. Transf. **133**, 364 (2014).

[28] K. Sasihithlu and A. Narayanaswamy, Phys. Rev. B **83**, 161406 (2011).

[29] C. Otey and S. Fan, Phys. Rev. B **84**, 245431 (2011).

[30] V.A. Golyk, M. Krüger, A.P. McCauley, and M. Kardar, EPL (Europhysics Lett. **101**, 34002 (2013).

[31] K.L. Nguyen, O. Merchiers, and P.O. Chapuis, J. Quant. Spectrosc. Radiat. Transf. **202**, 154 (2017).

[32] S.A. Biehs, M. Tschikin, and P. Ben-Abdallah, Phys. Rev. Lett. **109**, 104301 (2012).

[33] S. Molesky, P.S. Venkataram, W. Jin, and A.W. Rodriguez, Phys. Rev. B **101**, 35408 (2020).

[34] P.S. Venkataram, S. Molesky, W. Jin, and A.W. Rodriguez, Phys. Rev. Lett. **124**, 13904 (2020).

[35] C.R. Otey, W.T. Lau, and S. Fan, Phys. Rev. Lett. **104**, 154301 (2010).

[36] P. Ben-Abdallah and S.A. Biehs, Phys. Rev. Lett. **112**, 044301 (2014).

[37] P.J. van Zwol, L. Ranno, and J. Chevrier, Phys. Rev. Lett. **108**, 234301 (2012).

[38] K. Ito, K. Nishikawa, H. Iizuka, and H. Toshiyoshi, Appl. Phys. Lett. **105**, 253503 (2014).

[39] F. Menges, M. Dittberner, L. Novotny, D. Passarello, S.S.P. Parkin, M. Spieser, H. Riel, and B. Gotsmann, Appl. Phys. Lett. **108**, 171904 (2016).

[40] A. Fiorino, D. Thompson, L. Zhu, R. Mittapally, S.-A. Biehs, O. Bezencenet, N. El-Bondry, S. Bansropun, P. Ben-Abdallah, E. Meyhofer, and P. Reddy, ACS Nano **12**, 5774 (2018).

[41] K. Ito, K. Nishikawa, A. Miura, H. Toshiyoshi, and H. Iizuka, Nano Lett. **17**, 4347 (2017).

[42] A. Ott, R. Messina, P. Ben-Abdallah, and S.-A. Biehs, J. Photonics Energy **9**, 1 (2019).

[43] P. Ben-Abdallah and S.A. Biehs, Zeitschrift Fur Naturforsch. - Sect. A J. Phys. Sci. **72**, 151 (2017).

[44] T. Inoue, T. Koyama, D.D. Kang, K. Ikeda, T. Asano, and S. Noda, Nano Lett. **19**, 3948 (2019).

[45] A. Babuty, K. Joulain, P.O. Chapuis, J.J. Greffet, and Y. De Wilde, Phys. Rev. Lett. **110**, 146103 (2013).

[46] C.M. Hargreaves, Phys. Lett. A **30**, 491 (1969).

[47] G.A. Domoto, R.F. Boehm, and C.L. Tien, J. Heat Transfer **92**, 412 (1970).

[48] P. Sabbaghi, L. Long, X. Ying, L. Lambert, S. Taylor, C. Messner, and L. Wang, J. Appl. Phys. **128**, 025305 (2020).

[49] L. Tang, J. DeSutter, and M. Francoeur, ACS Photonics **7**, 1304 (2020).

[50] S. Lang, G. Sharma, S. Molesky, P.U. Kränzien, T. Jalas, Z. Jacob, A.Y. Petrov, and M. Eich, Sci. Rep. **7**, (2017).

[51] L. Hu, A. Narayanaswamy, X. Chen, and G. Chen, Appl. Phys. Lett. **92**, 133106 (2008).

[52] R.S. Ottens, V. Quetschke, S. Wise, A.A. Alemi, R. Lundock, G. Mueller, D.H. Reitze, D.B. Tanner, and B.F. Whiting, Phys. Rev. Lett. **107**, 014301 (2011).

[53] T. Kralik, P. Hanzelka, V. Musilova, A. Srnka, and M. Zobac, Rev. Sci. Instrum. **82**, 55106 (2011).

[54] M. Lim, J. Song, S.S. Lee, and B.J. Lee, Nat. Commun. **9**, 4302 (2018).

[55] M. Ghashami, H. Geng, T. Kim, N. Iacopino, S.K. Cho, and K. Park, Phys. Rev. Lett. **120**, 175901 (2018).

[56] X. Ying, P. Sabbaghi, N. Sluder, and L. Wang, ACS Photonics **7**, 190 (2019).

[57] K. Shi, Y. Sun, Z. Chen, N. He, F. Bao, J. Evans, and S. He, Nano Lett. **19**, 8082 (2019).

[58] M. Lim, J. Song, S.S. Lee, J. Lee, and B.J. Lee, Phys. Rev. Appl. **14**, 14070 (2020).

[59] S. Basu, Z.M. Zhang, and C.J. Fu, Int. J. Energy Res. **33**, 1203 (2009).

[60] A. Fiorino, L. Zhu, D. Thompson, R. Mittapally, P. Reddy, and E. Meyhofer, Nat. Nanotechnol. **13**, 806 (2018).

[61] G.R. Bhatt, B. Zhao, S. Roberts, I. Datta, A. Mohanty, T. Lin, J.-M. Hartmann, R. St-Gelais, S. Fan, and M. Lipson, Nat. Commun. **11**, 2545 (2020).

[62] C. Lucchesi, D. Cakiroglu, J.-P. Perez, T. Taliercio, E. Tournié, P.-O. Chapuis, and R. Vaillon, ArXiv **1912.09394**, (2019).

[63] S. Basu, Y.-B. Chen, and Z.M. Zhang, Int. J. Energy Res. **31**, 689 (2007).

[64] A. Lenert, D.M. Bierman, Y. Nam, W.R. Chan, I. Celanović, M. Soljačić, and E.N. Wang, Nat. Nanotechnol. **9**, 126 (2014).

[65] M. Elzouka and S. Ndao, Sol. Energy **141**, 323 (2017).

[66] A. Datas, Appl. Phys. Lett. **108**, 143503 (2016).

[67] A. Datas and R. Vaillon, Nano Energy **61**, 10 (2019).

[68] A. Datas and R. Vaillon, Appl. Phys. Lett. **114**, 133501 (2019).

[69] L. Zhu, A. Fiorino, D. Thompson, R. Mittapally, E. Meyhofer, and P. Reddy, Nature **566**, 239 (2019).

[70] E. Tervo, E. Bagherisereshki, and Z. Zhang, Front. Energy **12**, 5 (2018).

[71] M.P. Bernardi, D. Milovich, and M. Francoeur, Nat. Commun. **7**, 12900 (2016).

[72] C. Feng, Z. Tang, J. Yu, and C. Sun, Sensors **13**, 1998 (2013).

[73] C. Feng, Z.-A. Tang, and J. Yu, Chinese Phys. Lett. **29**, 38502 (2012).

[74] H. Salihoglu, W. Nam, L. Traverso, M. Segovia, P.K. Venuthurumilli, W. Liu, Y. Wei, W. Li, and X. Xu, Nano Lett. **20**, 6091 (2020).

[75] R. St-Gelais, B. Guha, L. Zhu, S. Fan, and M. Lipson, Nano Lett. **14**, 6971 (2014).

[76] B. Song, Y. Ganjeh, S. Sadat, D. Thompson, A. Fiorino, V. Fernández-Hurtado, J. Feist, F.J. Garcia-Vidal, J.C. Cuevas, P. Reddy, and E. Meyhofer, Nat. Nanotechnol. **10**, 253 (2015).

[77] B. Song, D. Thompson, A. Fiorino, Y. Ganjeh, P. Reddy, and E. Meyhofer, Nat. Nanotechnol. **11**, 509 (2016).

[78] R. St-Gelais, L. Zhu, S. Fan, and M. Lipson, Nat. Nanotechnol. **11**, 515 (2016).

[79] A. Fiorino, D. Thompson, L. Zhu, B. Song, P. Reddy, and E. Meyhofer, Nano Lett. **18**, 3711 (2018).

[80] D. Thompson, L. Zhu, E. Meyhofer, and P. Reddy, Nat. Nanotechnol. **15**, 99 (2020).

[81] J.-B. Xu, K. Läuger, R. Möller, K. Dransfeld, and I.H. Wilson, J. Appl. Phys. **76**, 7209 (1994).

[82] A. Narayanaswamy, S. Shen, and G. Chen, Phys. Rev. B - Condens. Matter Mater. Phys. **78**, 115303 (2008).

[83] E. Rousseau, A. Siria, G. Jourdan, S. Volz, F. Comin, J. Chevrier, and J.J. Greffet, Nat. Photonics **3**, 514 (2009).

[84] S. Shen, A. Narayanaswamy, and G. Chen, Nano Lett. **9**, 2909 (2009).

[85] P.J. van Zwol, S. Thiele, C. Berger, W.A. de Heer, and J. Chevrier, Phys. Rev. Lett. **109**, 264301 (2012).

[86] S. Shen, A. Mavrokefalos, P. Sambegoro, and G. Chen, Appl. Phys. Lett. **100**, 233114 (2012).

[87] S. Edalatpour and M. Francoeur, J. Quant. Spectrosc. Radiat. Transf. **118**, 75 (2013).

[88] P.J. van Zwol, D.F. Vles, W.P. Voorthuijzen, M. Péter, H. Vermeulen, W.J. van der Zande, J.M. Sturm, R.W.E. van de Kruijs, and F. Bijkerk, J. Appl. Phys. **118**, 213107 (2015).

[89] W. Müller-Hirsch, A. Kraft, M.T. Hirsch, J. Parisi, and A. Kittel, J. Vac. Sci. Technol. A Vacuum, Surfaces, Film. **17**, 1205 (1999).

[90] A. Kittel, W. Müller-Hirsch, J. Parisi, S.A. Biehs, D. Reddig, and M. Holthaus, Phys. Rev. Lett. **95**, 224301 (2005).

[91] L. Worbes, D. Hellmann, and A. Kittel, Phys. Rev. Lett. **110**, 134302 (2013).

[92] K. Kim, B. Song, V. Fernández-Hurtado, W. Lee, W. Jeong, L. Cui, D. Thompson, J. Feist, M.T.H. Reid, F.J. García-Vidal, J.C. Cuevas, E. Meyhofer, and P. Reddy, Nature **528**, 387 (2015).

[93] K. Kloppstech, N. Könne, S.A. Biehs, A.W. Rodriguez, L. Worbes, D. Hellmann, and A. Kittel, Nat. Commun. **8**, 14475 (2017).

[94] L. Cui, W. Jeong, V. Fernández-Hurtado, J. Feist, F.J. García-Vidal, J.C. Cuevas, E. Meyhofer, and P. Reddy, Nat. Commun. **8**, 14479 (2017).

[95] K. Kim, W. Jeong, W. Lee, and P. Reddy, ACS Nano **6**, 4248 (2012).

[96] J. Shi, B. Liu, P. Li, L.Y. Ng, and S. Shen, Nano Lett. **15**, 1217 (2015).

[97] F. Giazotto, T.T. Heikkilä, A. Luukanen, A.M. Savin, and J.P. Pekola, Rev. Mod. Phys. **78**, 217 (2006).

[98] Y. De Wilde, F. Formanek, R. Carminati, B. Gralak, P.-A. Lemoine, K. Joulain, J.-P. Mulet, Y. Chen, and J.-J. Greffet, Nature **444**, 740 (2006).

[99] A.C. Jones and M.B. Raschke, Nano Lett. **12**, 1475 (2012).

[100] Y. Kajihara, K. Kosaka, and S. Komiyama, Rev. Sci. Instrum. **81**, 033706 (2010).

[101] F. Huth, M. Schnell, J. Wittborn, N. Ocelic, and R. Hillenbrand, Nat. Mater. **10**, 352 (2011).

[102] Q. Weng, S. Komiyama, L. Yang, Z. An, P. Chen, S.-A. Biehs, Y. Kajihara, and W. Lu, Science (80-. ). **360**, 775 (2018).

[103] A. Kittel, U.F. Wischnath, J. Welker, O. Huth, F. Rüting, and S.-A. Biehs, Appl. Phys. Lett. **93**, 193109 (2008).

[104] P.O. Chapuis, S. Volz, C. Henkel, K. Joulain, and J.J. Greffet, Phys. Rev. B **77**, 035431 (2008).

[105] F. Singer, Y. Ezzahri, and K. Joulain, J. Quant. Spectrosc. Radiat. Transf. **154**, 55 (2015).

[106] W. Li and S. Fan, Opt. Express **26**, 15995 (2018).

[107] T. Ijiro and N. Yamada, Appl. Phys. Lett. **106**, 23103 (2015).

[108] D.-Z.A. Chen, A. Narayanaswamy, and G. Chen, Phys. Rev. B **72**, 155435 (2005).

[109] E.J. Tervo, O.S. Adewuyi, J.S. Hammonds, and B.A. Cola, Mater. Horiz. **3**, 434 (2016).

[110] X. Liu and Z. Zhang, ACS Photonics **2**, 1320 (2015).

[111] O. Quevedo-Teruel, H. Chen, A. Díaz-Rubio, G. Gok, A. Grbic, G. Minatti, E. Martini, S. Maci, G. V Eleftheriades, M. Chen, N.I. Zheludev, N. Papasimakis, S. Choudhury, Z.A. Kudyshev, S. Saha, H. Reddy, A. Boltasseva, V.M. Shalaev, A. V Kildishev, D. Sievenpiper, C. Caloz, A. Alù, Q. He, L. Zhou, G. Valerio, E. Rajo-Iglesias, Z. Sipus, F. Mesa, R. Rodríguez-Berral, F. Medina, V. Asadchy, S. Tretyakov, and C. Craeye, J. Opt. **21**, 73002 (2019).

[112] J. Yang, W. Du, Y. Su, Y. Fu, S. Gong, S. He, and Y. Ma, Nat.

Commun. **9**, 4033 (2018).
[113] N.H. Thomas, M.C. Sherrott, J. Broulliet, H.A. Atwater, and A.J. Minnich, Nano Lett. **19**, 3898 (2019).
[114] A.I. Volokitin and B.N.J. Persson, Phys. Rev. B **69**, 45417 (2004).
[115] C. Henkel and M. Wilkens, Eur. Lett. **47**, 414 (1999).
[116] W.E. Moerner, Y. Shechtman, and Q. Wang, Faraday Discuss. **184**, 9 (2015).